\documentclass[conference]{IEEEtran}
\IEEEoverridecommandlockouts
\usepackage{cite}
\usepackage{amsmath,amssymb,amsfonts}
\usepackage{algorithmic}
\usepackage{graphicx}
\usepackage{textcomp}
\usepackage{xcolor}

\usepackage{subcaption}
\usepackage{graphicx}
\usepackage{mathtools}
\usepackage{amsmath}
\usepackage{gensymb}
\usepackage{booktabs}

\usepackage{tasks}
\usepackage{multirow}
\usepackage{graphicx}
\usepackage{caption}
\usepackage{enumitem}

\usepackage{balance}

\usepackage{pgfplots}
\pgfplotsset{compat=1.18}

\usepackage{fancyhdr}

\usepackage[font=footnotesize,labelformat=simple]{subcaption}

\def\BibTeX{{\rm B\kern-.05em{\sc i\kern-.025em b}\kern-.08em
    T\kern-.1667em\lower.7ex\hbox{E}\kern-.125emX}}
\begin{document}

\makeatletter

\def\ps@IEEEtitlepagestyle{%
    \def\@oddfoot{\mycopyrightnotice}%
    \def\@evenfoot{}%
}
\def\mycopyrightnotice{%
}


\makeatletter
 \let\old@ps@IEEEtitlepagestyle\ps@IEEEtitlepagestyle
\def\confheader#1{%
    \def\ps@IEEEtitlepagestyle{%
        \old@ps@IEEEtitlepagestyle%
        \def\@oddhead{\strut\hfill#1\hfill\strut}%
        \def\@evenhead{\strut\hfill#1\hfill\strut}%
    }%
    \ps@headings%
}
 \makeatother

\confheader{%
}


\makeatletter
\newcommand{\algrule}[1][.2pt]{\par\vskip.5\baselineskip\hrule height #1\par\vskip.5\baselineskip}
\makeatother

\thispagestyle{plain}
\pagestyle{plain}

\title{An IoT Based Smart Waste Management System for the Municipality or City Corporations}

\author{
\IEEEauthorblockN{Laboni Paul\textsuperscript{1}, Rahul Deb Mohalder\textsuperscript{2}, Kazi Masudul Alam\textsuperscript{3}}
    \IEEEauthorblockA{\textsuperscript{1}\textit{VISIE Ltd., Dhaka, Bangladesh} }
    \IEEEauthorblockA{\textsuperscript{2}\textit{ICT Cell, Khulna University, Khulna, Bangladesh} }
    \IEEEauthorblockA{\textsuperscript{3}\textit{Computer Science and Engineering Discipline, Khulna University, Khulna, Bangladesh} } 
    \IEEEauthorblockA{Email- laboni1124@cseku.ac.bd, rahul@ku.ac.bd, kazi@cse.ku.ac.bd}

}

\maketitle

\begin{abstract}
The population of the urban areas is increasing daily, and this migration is causing serious environmental pollution. A larger population is creating pressure on the municipality's waste management and the city corporations of developing countries such as Bangladesh, further threatening human health. New generation technologies, such as the Internet of Things (IoT)-based waste management systems, can help improve this serious issue. IoT-enabled smart dustbins and mobile applications-based interactive management can effectively solve this problem. In this article, we combine these two technologies to offer an acceptable solution to this problem. The proposed waste management model enables smart dustbins to communicate with waste collectors or waste control centers whenever it is necessary. Additionally, city dwellers can use mobile applications to report their observations in their neighborhoods. As a result, both sensors and humans are involved directly in the development loop. We have conducted a detailed survey to study the acceptance of such a system in the community and received some encouraging results. 
\end{abstract}


\section{Introduction}
\label{sec_intro}
The waste that the tumid population produces is primarily to blame for the environmental pollution of the urban environment. Food waste is one of them. It is increasing the environmental and health risk of urban people \cite{logan2019investigating, wen2018design}. We have to spend a huge amount of money on removing this urban waste every year. It has become difficult to remove this excessive waste from people's unconsciousness. For this, experts are looking for a smart solution to remove waste. From the World Bank's report every year 1.3 billion tons of household waste is generated. And it will be 2.2 billion tons in 2025 \cite{hoornweg2013environment}. Our era is technology-based and at present, the internet is very common in Bangladesh. The price of the internet is decreasing day by day and the use of the internet is much easier than before. So, an interconnected system will be a good solution for a waste management system. Interconnected systems may decrease cost and time. This will make waste management more efficient and convenient.

For an increasing number of people, extra waste is being produced. But waste management system capability has not increased for this growing population. As compared to the population, our waste management system manpower is very low and there is a shortage of garbage bins and vehicles. As a developing country, Bangladesh can provide a minimum amount of money for the waste management system. But most of them are used for buying vehicles or paying the waste collector. So our waste management department manages some open space garbage locations. It is very unhygienic as most of the waste of our country is rotting garbage. As most of the garbage bins are open so easily it pollutes the air. So a large number of people are suffering from air pollution diseases. So we need a proper and efficient waste management system.

Now we are living in the era of IoT. A great revelation has occurred in the industrial and residential sectors through IoT service. People are moving to IoT devices \cite{vorakulpipat2018recent}. IoT-based engineering and scientific application is also involved \cite{alavi2018internet}. Smart city-based applications are one of them. There is no alternate smart city-based IoT application for giving services and monitoring an upgrowing population \cite{silva2018towards}.

\begin{figure}
	\centering
	\includegraphics[height=0.35\textwidth, width=0.48\textwidth, scale=0.5]{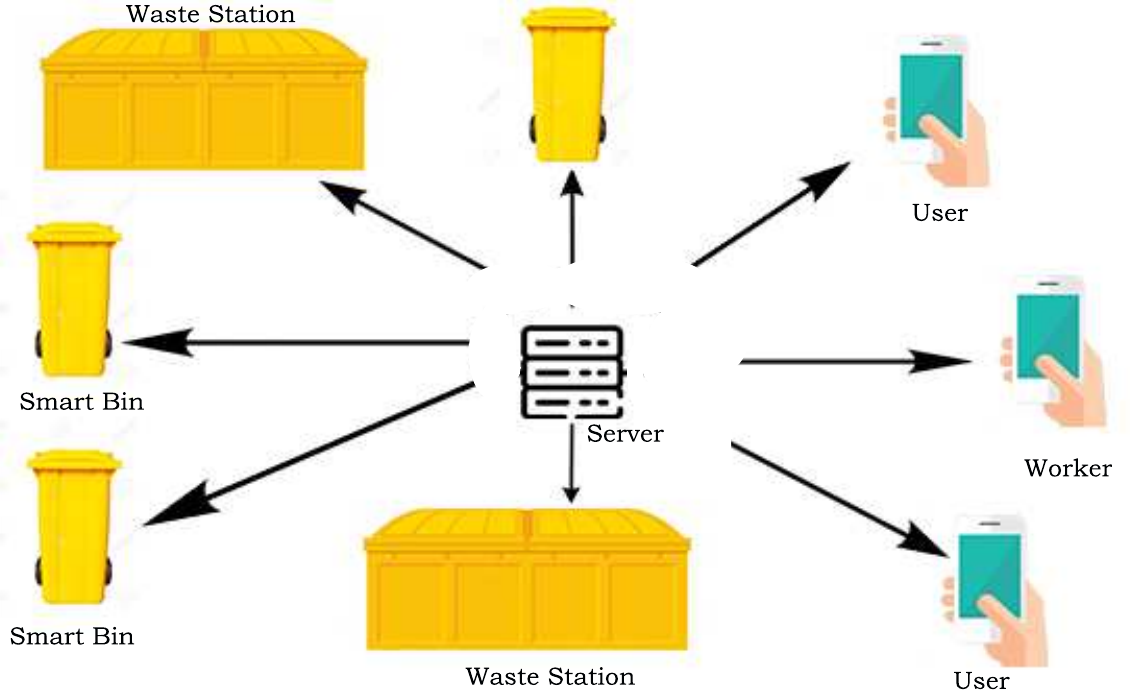}
	\caption{Smart waste management system}
	\label{fig:smart_waste}
\end{figure}

Governments of every country are trying to make everything digitized and smart through IoT services. Because IoT is a trendy technology in recent times \cite{kumar2019recent}. IoT-based monitoring systems have become popular all over the world. Not only for home or office management systems but it is also used broadly in development for environmental monitoring systems \cite{geng2019mobile, marques2018system}. But our South Asian countries are still backward. In every country, the Internet is available in any place and cheaper than before. And today most of people are using the internet \cite{mohalder2019iot}. So, we made a smart system for waste management using IoT, which can be easily implemented. Our system may be a bit costly but it is very sustainable and efficient. And it can lead any country's next vision SMART CITY one step ahead (Figure \ref{fig:smart_waste}).


\begin{figure}
	\centering
	\includegraphics[height=0.73\textwidth, width=0.45\textwidth, scale=0.5]{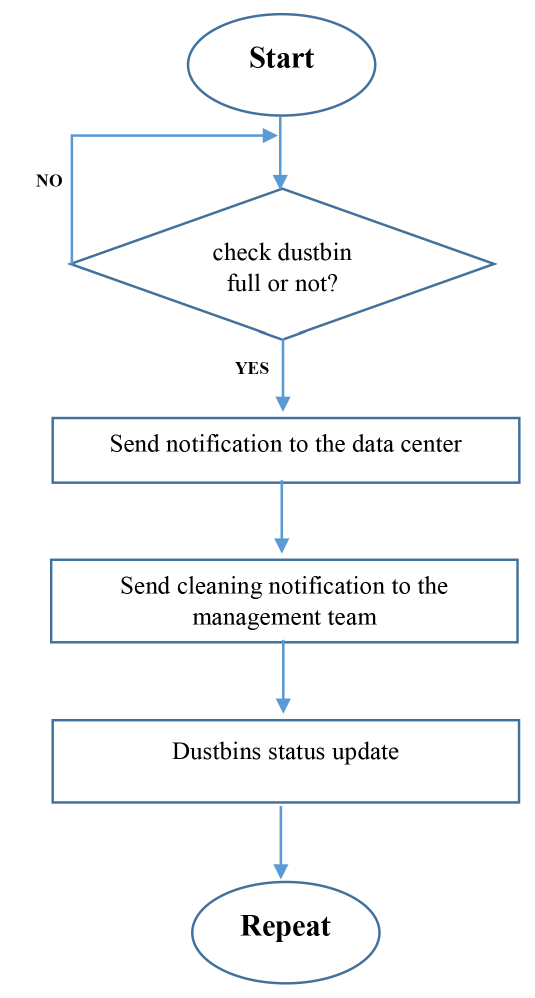}
	\caption{Working flow chart of a smart dustbin.}
	\label{fig:proposed_system_architecture}
\end{figure}

In this paper, we present a smart waste management model for the municipality and the city corporations where IoT-enabled sensors collect garbage status from the dustbins, and city dwellers can also participate in improving their neighborhood status. The rest of the paper is organized as related works in section II, proposed system architecture in section III,  system implementation and experimental analysis in section IV, and the conclusion in the final section.

\section{Related Works}
Smart garbage and waste collection systems are designed with a smart bin and a web page Wi-Fi enabled. Using a micro-controller-based system smart bin is designed which includes an IR wireless system and it is connected with the central system. All updated status is shown to the web page by it and it is uploaded through a Wi-Fi network. By smartphone, anyone can access data from a web page \cite{navghane2016iot}.

In \cite{folianto2015smartbin}, here presented a smart bin system to identify the fullness of litter bins. This system collects data from the litter bin and delivers the data to the cleaner using a wireless network (mesh network). It also reduces employee's workload and stops extra time-consuming tasks. When data is sent from smart bins it is received by the data receiving module and this module stores the data in the database. By using the proxy module of the bin sub-system information can be accessed from the workstation.

This SMS-based solid waste management system. To design this system here used a combination of Zigbee and GSM technologies. An ARM 7 controller is used in every bin. When the garbage reaches the sensor level the ARM 7 controllers get a signal. Then the controller sends an SMS to the garbage or waste-collecting person of the garbage collector vehicles to collect the garbage as soon as possible. This SMS indicates that the bins are almost filled with garbage. To send SMS, the ARM 7 controller uses GSM technology \cite{mahajan2014waste}.  

Smart bin application based on information self-contained in tags associated with each waste item. When any waste is thrown into the smart bin then the smart bin can track the waste by using RFID RFID-based system, and for this, it needs no support from any external information system. There are two important features of this system. Firstly, the application will be very helpful to the user for properly sorting waste. Secondly, every smart bin knows what it contains and can report to the authority for recycling \cite{glouche2013smart}. In \cite{zahur2019iot}, IoT and cloud-based frameworks were used for waste management. There was a small smart bin and associated application in their system.  

We have designed and developed an IoT-based waste management system to properly manage civic waste and motivate citizens to fulfill their civic responsibilities by raising awareness among city dwellers. Our system is more convenient and user-friendly than others. Through our system, municipal or corporation workers can manage their day-to-day work properly, city dwellers can report their problems, and dustbin overflows can be stopped.

\section{Proposed System Architecture}
This system consists of two parts. One part is the hardware base and another is the mobile application base. In the hardware part “Smart Dustbin” has been made using a microcontroller, internet-enabled system, and sensors (Figure \ref{fig:proposed_system_architecture}). We have categorized dustbins into two sectors. One is a waste station in a municipal area where big dustbins are used for storing collected residence area waste, and another is small smart dustbins that are placed on the roadside, in front of residences, offices, schools/colleges, shops, markets, and another place.

\begin{figure}
	\centering
	\includegraphics[height=0.42\textwidth, width=0.48\textwidth, scale=0.5]{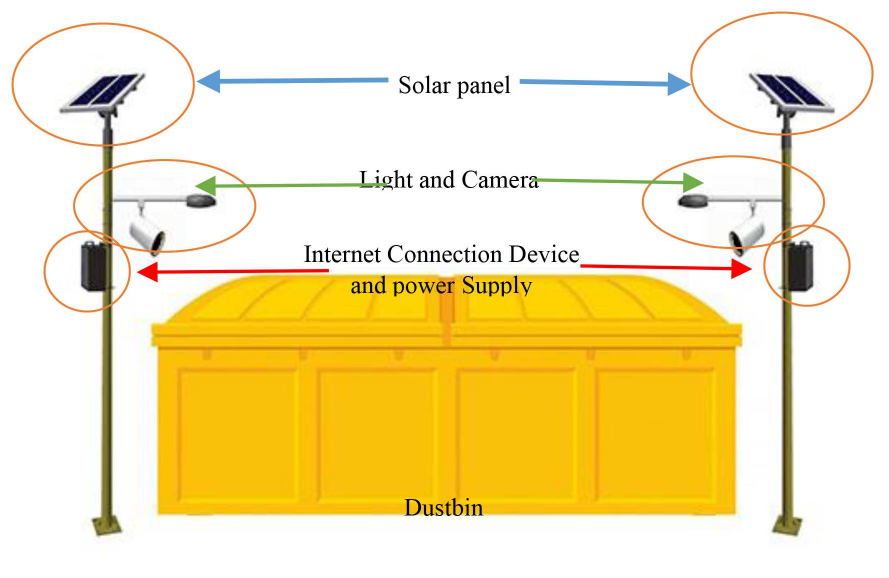}
	\caption{Larger dustbins (waste station).}
	\label{fig:proposed_smart_dustbin}
\end{figure}

In the waste station, we set two solar light posts including a camera in the left and right direction of the station (Figure \ref{fig:proposed_smart_dustbin}). We set the camera in the light post for monitoring dustbins. Every 10 or 15 minutes the captured image from the camera is sent to the central data center of the city corporation waste management system using a Wi-Fi network. After processing the imaging system like Sarc et al. \cite{sarc2019digitalisation}, the system decides whether the dustbin is full, overflow, or empty. If it’s full or overflowing then the system sends a notification or SMS including the location of the waste station to the waste management team for an attempt as soon as possible. Using a smartphone workers can find the waste station Google Map location from the notification. But if he has not, he can get the exact location address of the waste station. For maintaining the camera we used solar light system power and solar light for capturing images at night. Because the camera's night vision image is not sufficient for making decisions after processing that image.

In small dustbins, ultrasonic sensors have been used like Zahur et al.\cite{zahur2019iot}. Using ultrasonic sensors, waste height is measured inside the dustbin. Ultrasonic Sensors work based on ultrasound waves. It sends ultrasonic sounds in specific directions. When it gets a barrier, its ultrasound waves return to the sensor. They used a Temperature Sensor to sense the Smart Dustbin’s inner temperature. Dustbin’s inner temperature can give us information about the wastage type. Three types of LED light (Green, Yellow, and Red) have been added so that anyone can understand whether the dustbin is full or empty. Green light indicates the dustbin is empty, Yellow light indicates the dustbin is full above 50\% of its height and Red light turns on when the dustbin is full. By Fog Computing dustbin full signal has been collected every 10 or 15 minutes from the surrounding dustbins of a network zone and sent to the central data center of the city corporation’s fecal waste management system. For sending data to the data center and surrounding area Wi-Fi facilities are used. Public Wi-Fi network zones (set by the City Corporation authority) and personal (office, residence, shop) Wi-Fi network zones are available in our city.

But for the remote areas where Wi-Fi network facilities are not available, we used the GSM/GPRS module for sending data. When the system gets the dustbin full signal then it communicates with the waste management term via smartphone notification or mobile SMS. For the power supply, we added a Solar panel to our system. So that it can run without any external or grid power.

\begin{figure*} 
	\centering
	\includegraphics[height=0.68\textwidth, width=0.65\textwidth, scale=0.5]{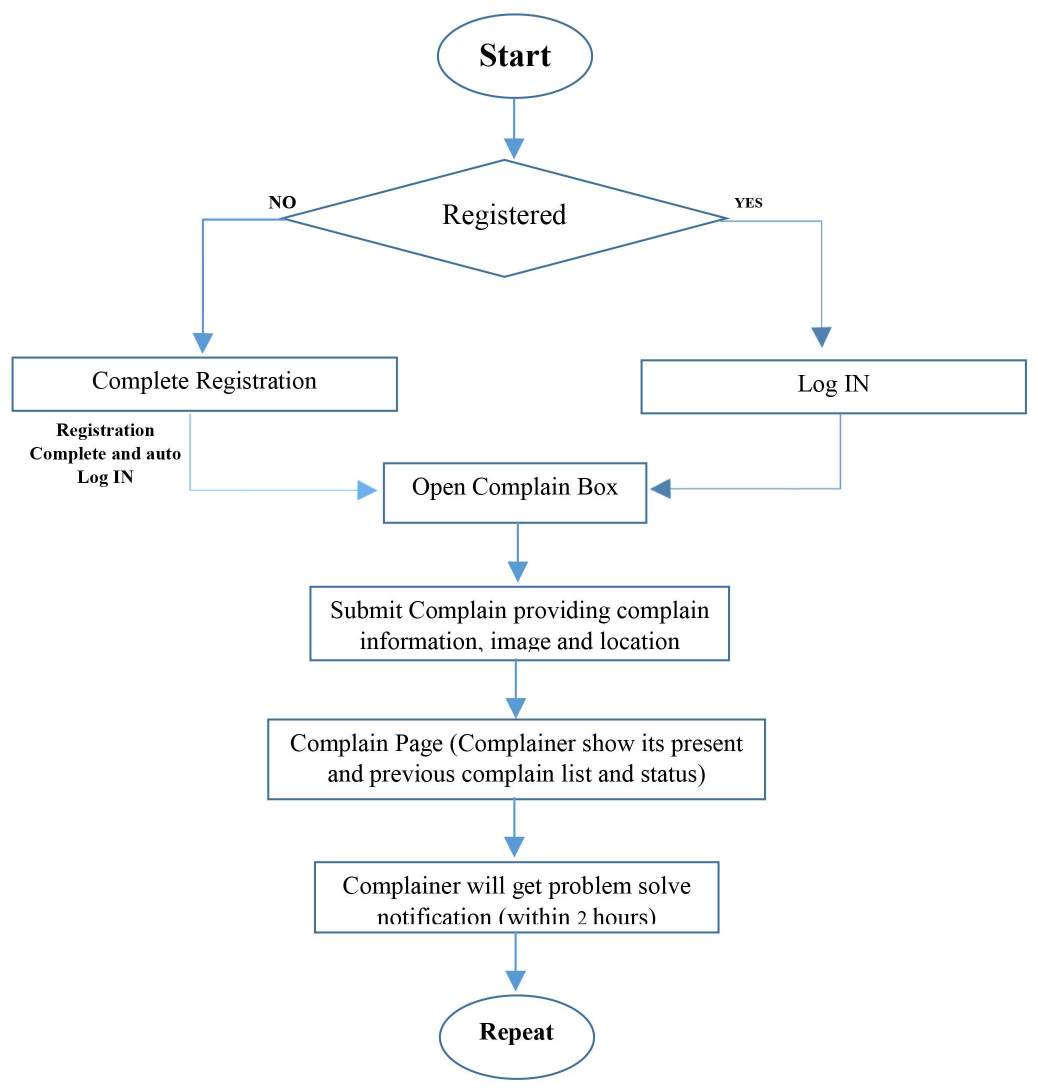}
	\caption{Mobile application city dweller interaction model}
	\label{fig:proposed_mobile_app_system_architecture}
\end{figure*}

Citizens in the municipal area can use our Mobile Application system and complain about keeping our city clean (Figure \ref{fig:proposed_mobile_app_system_architecture}). Using their National Identification (NID) card number and other information they registered in the system. When any citizen notices that waste is kept on the roadside, school/college, residence area, office, or any place but no one cleans it and the waste pollutes the air then he/she can complain to the waste management system using the application.


\begin{figure}
	\centering
	\includegraphics[height=0.5\textwidth, width=0.47\textwidth, scale=0.5]{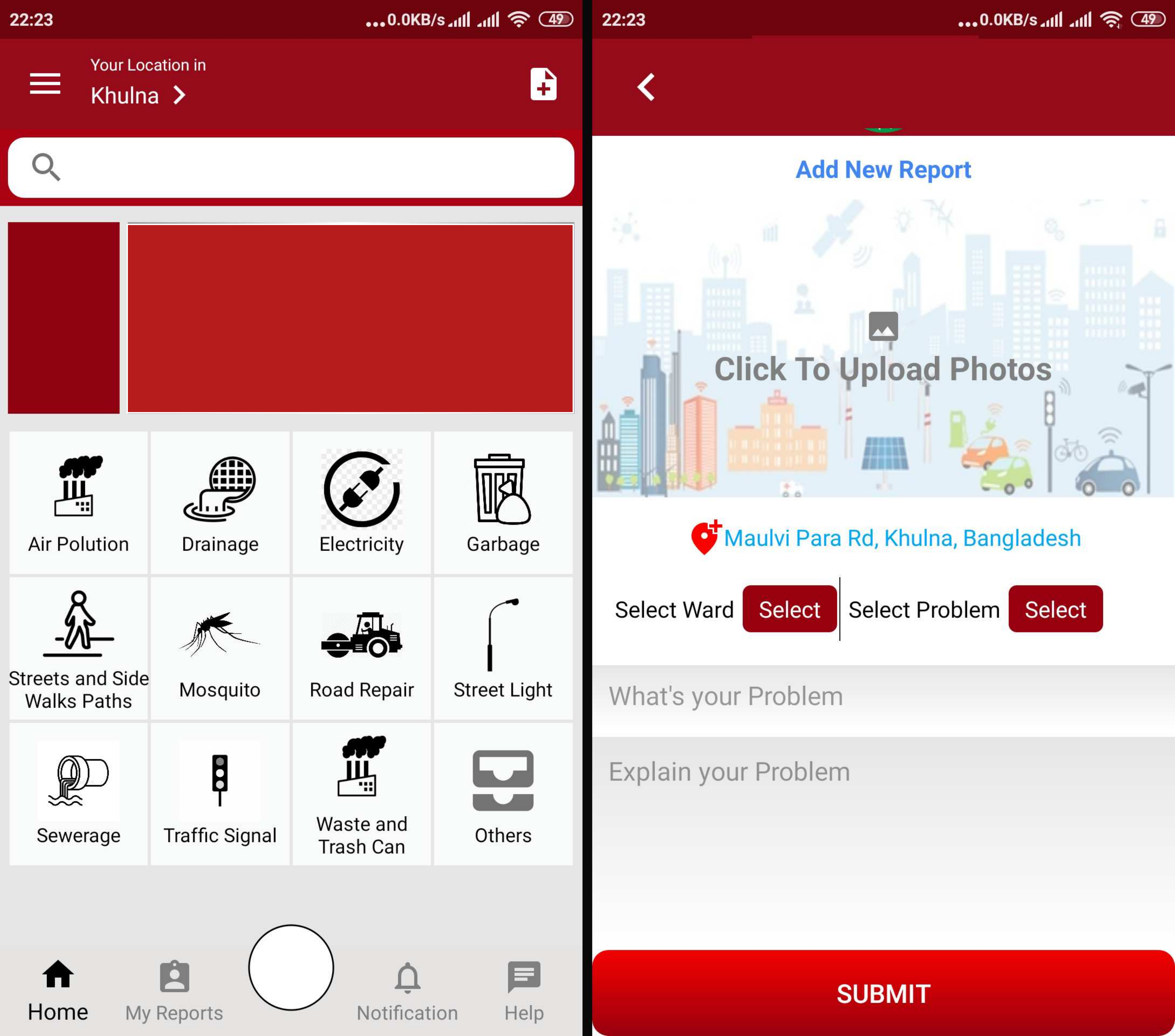}
	\caption{Screen of the developed mobile application}
	\label{fig:apps_ui}
\end{figure}

By using this application he/she takes an image of the place and submits the complaint. In complaint submitting time, the location of the place is tracked automatically by the application. If the location address is not correct then the complainant can edit the location. After getting the complaint our system sends a notification or SMS to the waste management team and they take action to solve that complaint. All complaints are solved within the next two or three hours. After solving that, our waste management team gives solved feedback from their mobile application, and the citizens who complain get a notification that “Your complaint has been solved. Thanks for your activity”. We use the NID card number for the registration process because no one can post fake complaints and if any we can identify that person easily (Figure \ref{fig:apps_ui}).

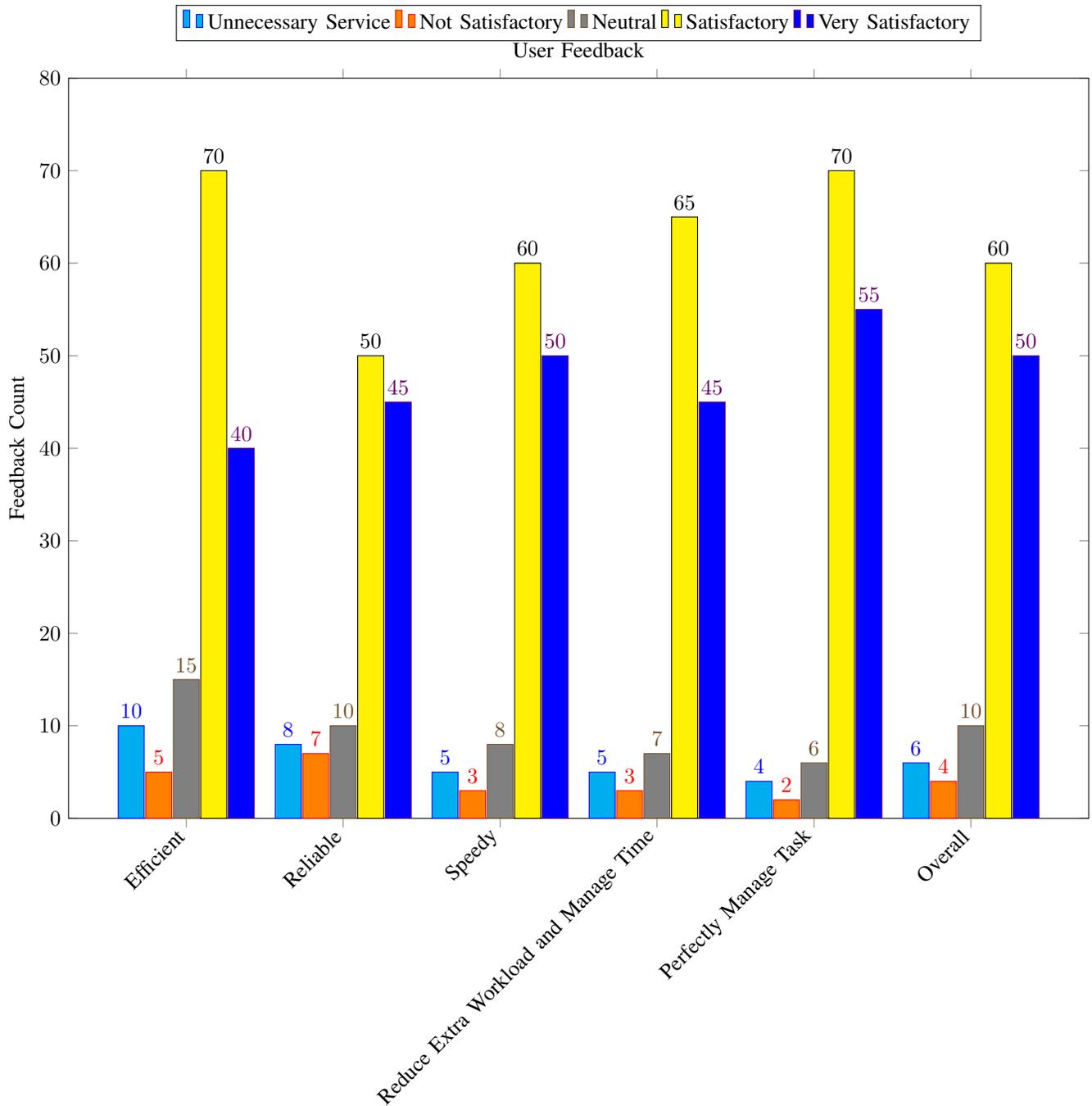
\begin{figure*}[h!]
    \centering
    \begin{tikzpicture}
        \begin{axis}[
            width=1.0\textwidth,
            height=0.75\textwidth,
            ybar=0.7pt,
            bar width=12pt,
            enlarge x limits=0.15,
            ymin=0, ymax=80,
            ylabel={Feedback Count},
            symbolic x coords={Efficient, Reliable, Speedy, Reduce Extra Workload and Manage Time, Perfectly Manage Task, Overall},
            xtick=data,
            nodes near coords,
            nodes near coords align={vertical},
            xticklabel style={rotate=45, anchor=east},
            legend style={at={(0.5,1.05)}, anchor=south, legend columns=-1},
            title={User Feedback}
        ]
        
        \addplot+[ybar, fill=cyan] plot coordinates {
            (Efficient, 10) (Reliable, 8) (Speedy, 5) (Reduce Extra Workload and Manage Time, 5) (Perfectly Manage Task, 4) (Overall, 6)
        };
        
        \addplot+[ybar, fill=orange] plot coordinates {
            (Efficient, 5) (Reliable, 7) (Speedy, 3) (Reduce Extra Workload and Manage Time, 3) (Perfectly Manage Task, 2) (Overall, 4)
        };
        
        \addplot+[ybar, fill=gray] plot coordinates {
            (Efficient, 15) (Reliable, 10) (Speedy, 8) (Reduce Extra Workload and Manage Time, 7) (Perfectly Manage Task, 6) (Overall, 10)
        };
        
        \addplot+[ybar, fill=yellow] plot coordinates {
            (Efficient, 70) (Reliable, 50) (Speedy, 60) (Reduce Extra Workload and Manage Time, 65) (Perfectly Manage Task, 70) (Overall, 60)
        };
        
        \addplot+[ybar, fill=blue] plot coordinates {
            (Efficient, 40) (Reliable, 45) (Speedy, 50) (Reduce Extra Workload and Manage Time, 45) (Perfectly Manage Task, 55) (Overall, 50)
        };
        
        \legend{Unnecessary Service, Not Satisfactory, Neutral, Satisfactory, Very Satisfactory}
        
        \end{axis}
    \end{tikzpicture}
    \caption{Usability analysis based on user feedback using Likert scale}
	\label{fig:user_feedback}
\end{figure*}


\section{System Implementation and Experimental Analysis}
Our system has two parts, one is the hardware part and another is the software part. The hardware part consists of hardware tools, circuits, and the software part consists of mobile applications.

\subsection{Hardware Part}
For implementing our system hardware part we used these tools \textit{ Arduino Uno R3, SIM 900A (GSM with GPRS module), DYP-ME007 V2 Ultrasonic Sensor, NEO 6M GPS module,  9 Amp, 12V Battery, 10W Solar Panel, Camera, Router, Internet Connection}.



\subsection{Software Part}
 The software part consists of two parts. One is the web application and another part is the mobile application. In the web application part, there is an admin panel and a web server. We developed the admin site using \textit{laravel} framework and used \textit{mySQL} for the server. We have developed Android and iOS mobile applications using \textit{Android Studio} and \textit{Xcode}. 

\subsection{Performance Analysis}
Our waste management systems mobile application (android and iOS) works smoothly on any updated and lower version OS and we tested in different lower and updated version smartphones. We used \textit{MySQL} server for the online database and used \textit{SQLite} for the app's internal database. Our mobile apps Google map responds so fast and accurately to shows or tracks the current or desired locations. We set firebase cloud messaging (FCM) service in our apps and web service to communicate with users and admin. FCM service is faster than any other service. It delivers any message or notification as soon as possible. Our web portal is suitable for Google Chrome, Mozilla Firefox, and edge browsers. It works smoothly without any interruption. Our smart dustbin system response and monitoring system are so fast and accurate. It sends a Lisbon monitoring report so fast when it becomes full and empty.

\subsection{Usability Analysis}
We used some questionnaires to analyze our system usability and completed a short survey with 150 people. Those questions were whether our system is efficient, reliable, speedy”, reduces extra workload and manages time, perfectly manages time and overall. We also categorized user feedback into Likert scales \cite{joshi2015likert}. Those are unnecessary service, not satisfactory, neutral, satisfactory, and very satisfactory. Figure \ref{fig:user_feedback} shows the user feedback of our survey report. From our survey data, we got 82\% positive feedback, 10\% no comment,  can, and 8\% negative feedback. From our survey report, we can say that our system can be a useful and better replacement for traditional waste management systems.

\section{Conclusion}
Smart waste management is an extremely challenging task. Because most people are not aware of their surroundings. Our IoT-based smart management system will help improve the city's waste management system and make the unaware city dwellers aware. We have developed a combination of Internet and mobile application-based systems because most people now rely on the Internet and smartphones, making our waste management system simple to use for any city dweller. The main limitations of our system are the speed of the Internet and electricity availability. In the future, we want to improve our system it can run with low electrical power and internet speed.


\bibliographystyle{IEEEtran}
\bibliography{reference}

\begin{thebibliography}{10}
\providecommand{\url}[1]{#1}
\csname url@samestyle\endcsname
\providecommand{\newblock}{\relax}
\providecommand{\bibinfo}[2]{#2}
\providecommand{\BIBentrySTDinterwordspacing}{\spaceskip=0pt\relax}
\providecommand{\BIBentryALTinterwordstretchfactor}{4}
\providecommand{\BIBentryALTinterwordspacing}{\spaceskip=\fontdimen2\font plus
\BIBentryALTinterwordstretchfactor\fontdimen3\font minus \fontdimen4\font\relax}
\providecommand{\BIBforeignlanguage}[2]{{%
\expandafter\ifx\csname l@#1\endcsname\relax
\typeout{** WARNING: IEEEtran.bst: No hyphenation pattern has been}%
\typeout{** loaded for the language `#1'. Using the pattern for}%
\typeout{** the default language instead.}%
\else
\language=\csname l@#1\endcsname
\fi
#2}}
\providecommand{\BIBdecl}{\relax}
\BIBdecl

\bibitem{logan2019investigating}
M.~Logan, M.~Safi, P.~Lens, and C.~Visvanathan, ``Investigating the performance of internet of things based anaerobic digestion of food waste,'' \emph{Process safety and environmental protection}, vol. 127, pp. 277--287, 2019.

\bibitem{wen2018design}
Z.~Wen, S.~Hu, D.~De~Clercq, M.~B. Beck, H.~Zhang, H.~Zhang, F.~Fei, and J.~Liu, ``Design, implementation, and evaluation of an internet of things (iot) network system for restaurant food waste management,'' \emph{Waste management}, vol.~73, pp. 26--38, 2018.

\bibitem{hoornweg2013environment}
D.~Hoornweg, P.~Bhada-Tata, and C.~Kennedy, ``Environment: Waste production must peak this century,'' \emph{Nature News}, vol. 502, no. 7473, p. 615, 2013.

\bibitem{vorakulpipat2018recent}
C.~Vorakulpipat, E.~Rattanalerdnusorn, P.~Thaenkaew, and H.~D. Hai, ``Recent challenges, trends, and concerns related to iot security: An evolutionary study,'' in \emph{2018 20th International Conference on Advanced Communication Technology (ICACT)}.\hskip 1em plus 0.5em minus 0.4em\relax IEEE, 2018, pp. 405--410.

\bibitem{alavi2018internet}
A.~H. Alavi, P.~Jiao, W.~G. Buttlar, and N.~Lajnef, ``Internet of things-enabled smart cities: State-of-the-art and future trends,'' \emph{Measurement}, vol. 129, pp. 589--606, 2018.

\bibitem{silva2018towards}
B.~N. Silva, M.~Khan, and K.~Han, ``Towards sustainable smart cities: A review of trends, architectures, components, and open challenges in smart cities,'' \emph{Sustainable Cities and Society}, vol.~38, pp. 697--713, 2018.

\bibitem{kumar2019recent}
A.~Kumar, A.~O. Salau, S.~Gupta, and K.~Paliwal, ``Recent trends in iot and its requisition with iot built engineering: A review,'' \emph{Advances in Signal Processing and Communication}, pp. 15--25, 2019.

\bibitem{geng2019mobile}
X.~Geng, Q.~Zhang, Q.~Wei, T.~Zhang, Y.~Cai, Y.~Liang, and X.~Sun, ``A mobile greenhouse environment monitoring system based on the internet of things,'' \emph{IEEE Access}, vol.~7, pp. 135\,832--135\,844, 2019.

\bibitem{marques2018system}
G.~Marques, C.~Roque~Ferreira, and R.~Pitarma, ``A system based on the internet of things for real-time particle monitoring in buildings,'' \emph{International journal of environmental research and public health}, vol.~15, no.~4, p. 821, 2018.

\bibitem{mohalder2019iot}
R.~D. Mohalder, M.~A. Rahman, and A.~Saha, ``An iot based approach against physical and mental assault in educational institution,'' in \emph{2019 10th International Conference on Computing, Communication and Networking Technologies (ICCCNT)}, 2019, pp. 1--5.

\bibitem{navghane2016iot}
S.~Navghane, M.~Killedar, and V.~Rohokale, ``Iot based smart garbage and waste collection bin,'' \emph{International Journal of Advanced Research in Electronics and Communication Engineering (IJARECE)}, vol.~5, no.~5, pp. 1576--1578, 2016.

\bibitem{folianto2015smartbin}
F.~Folianto, Y.~S. Low, and W.~L. Yeow, ``Smartbin: Smart waste management system,'' in \emph{2015 IEEE Tenth International Conference on Intelligent Sensors, Sensor Networks and Information Processing (ISSNIP)}.\hskip 1em plus 0.5em minus 0.4em\relax IEEE, 2015, pp. 1--2.

\bibitem{mahajan2014waste}
K.~Mahajan and J.~Chitode, ``Waste bin monitoring system using integrated technologies,'' \emph{International Journal of Innovative Research in Science, Engineering and Technology}, vol.~3, no.~7, pp. 14\,953--14\,957, 2014.

\bibitem{glouche2013smart}
Y.~Glouche and P.~Couderc, ``A smart waste management with self-describing objects,'' in \emph{The Second International Conference on Smart Systems, Devices and Technologies (SMART'13)}, 2013.

\bibitem{zahur2019iot}
M.~I. Zahur, M.~S. Rahman, A.~Akther, and K.~M. Alam, ``An iot based green waste management system for bangladesh,'' in \emph{2019 1st International Conference on Advances in Science, Engineering and Robotics Technology (ICASERT)}.\hskip 1em plus 0.5em minus 0.4em\relax IEEE, 2019, pp. 1--6.

\bibitem{sarc2019digitalisation}
R.~Sarc, A.~Curtis, L.~Kandlbauer, K.~Khodier, K.~Lorber, and R.~Pomberger, ``Digitalisation and intelligent robotics in value chain of circular economy oriented waste management--a review,'' \emph{Waste Management}, vol.~95, pp. 476--492, 2019.

\bibitem{joshi2015likert}
A.~Joshi, S.~Kale, S.~Chandel, and D.~K. Pal, ``Likert scale: Explored and explained,'' \emph{British Journal of Applied Science \& Technology}, vol.~7, no.~4, p. 396, 2015.

\end{thebibliography}

\end{document}